\documentclass[aps,prl,twocolumn,showpacs,groupedaddress,letterpaper,fleqn]{revtex4}  
\usepackage{graphicx}  
\usepackage{dcolumn}   
\usepackage{bm}        
\usepackage{amssymb}   
\usepackage{amsmath}   
\usepackage{fleqn}

\newcommand{\Bbar}{\,\overline{\!B}}

\newcommand{\bsb}{\ensuremath{\Bbar{}_s}}
\newcommand{\bs}{\ensuremath{B_s}}

\newcommand{\etal}{\emph{et al.}}
\newcommand{\rf}[1]{Ref.\ \cite{#1}}

\newcommand{\asl} {A^{s,unt}_{SL}}
\newcommand{\asld} {A^{d,unt}_{SL}}

\newcommand{\dzero}{D0~}

\begin{document}



\hspace{5.2in} \mbox{Fermilab-Pub-07/005-E}

\title{Measurement of the Charge Asymmetry in Semileptonic $\bs$ Decays}
%
\author{                                                                      
V.M.~Abazov,$^{35}$                                                           
B.~Abbott,$^{75}$                                                             
M.~Abolins,$^{65}$                                                            
B.S.~Acharya,$^{28}$                                                          
M.~Adams,$^{51}$                                                              
T.~Adams,$^{49}$                                                              
E.~Aguilo,$^{5}$                                                              
S.H.~Ahn,$^{30}$                                                              
M.~Ahsan,$^{59}$                                                              
G.D.~Alexeev,$^{35}$
% ...
\\                                                                            
\vskip 0.30cm                                                                 
\centerline{(D0 Collaboration)}                                             
\vskip 0.30cm                                                                 
}                                                                             
\affiliation{                                                                 
\centerline{$^{1}$Universidad de Buenos Aires, Buenos Aires, Argentina}       
\centerline{$^{2}$LAFEX, Centro Brasileiro de Pesquisas F{\'\i}sicas,         
                  Rio de Janeiro, Brazil}                                     
\centerline{$^{3}$Universidade do Estado do Rio de Janeiro,                   
                  Rio de Janeiro, Brazil}                                     
\centerline{$^{4}$Instituto de F\'{\i}sica Te\'orica, Universidade            
                  Estadual Paulista, S\~ao Paulo, Brazil}                     
\centerline{$^{5}$University of Alberta, Edmonton, Alberta, Canada,           
                  Simon Fraser University, Burnaby, British Columbia, Canada,}
\centerline{York University, Toronto, Ontario, Canada, and                    
                  McGill University, Montreal, Quebec, Canada}                
\centerline{$^{6}$University of Science and Technology of China, Hefei,       
                  People's Republic of China}                                 
\centerline{$^{7}$Universidad de los Andes, Bogot\'{a}, Colombia}             
\centerline{$^{8}$Center for Particle Physics, Charles University,            
                  Prague, Czech Republic}                                     
\centerline{$^{9}$Czech Technical University, Prague, Czech Republic}         
\centerline{$^{10}$Center for Particle Physics, Institute of Physics,         
                   Academy of Sciences of the Czech Republic,                 
                   Prague, Czech Republic}                                    
\centerline{$^{11}$Universidad San Francisco de Quito, Quito, Ecuador}        
\centerline{$^{12}$Laboratoire de Physique Corpusculaire, IN2P3-CNRS,         
                   Universit\'e Blaise Pascal, Clermont-Ferrand, France}      
\centerline{$^{13}$Laboratoire de Physique Subatomique et de Cosmologie,      
                   IN2P3-CNRS, Universite de Grenoble 1, Grenoble, France}    
\centerline{$^{14}$CPPM, IN2P3-CNRS, Universit\'e de la M\'editerran\'ee,     
                   Marseille, France}                                         
\centerline{$^{15}$Laboratoire de l'Acc\'el\'erateur Lin\'eaire,              
                   IN2P3-CNRS et Universit\'e Paris-Sud, Orsay, France}       
\centerline{$^{16}$LPNHE, IN2P3-CNRS, Universit\'es Paris VI and VII,         
                   Paris, France}                                             
\centerline{$^{17}$DAPNIA/Service de Physique des Particules, CEA, Saclay,    
                   France}                                                    
\centerline{$^{18}$IPHC, IN2P3-CNRS, Universit\'e Louis Pasteur, Strasbourg,  
                   France, and Universit\'e de Haute Alsace,                  
                   Mulhouse, France}                                          
\centerline{$^{19}$Institut de Physique Nucl\'eaire de Lyon, IN2P3-CNRS,      
                   Universit\'e Claude Bernard, Villeurbanne, France}         
\centerline{$^{20}$III. Physikalisches Institut A, RWTH Aachen,               
                   Aachen, Germany}                                           
\centerline{$^{21}$Physikalisches Institut, Universit{\"a}t Bonn,             
                   Bonn, Germany}                                             
\centerline{$^{22}$Physikalisches Institut, Universit{\"a}t Freiburg,         
                   Freiburg, Germany}                                         
\centerline{$^{23}$Institut f{\"u}r Physik, Universit{\"a}t Mainz,            
                   Mainz, Germany}                                            
\centerline{$^{24}$Ludwig-Maximilians-Universit{\"a}t M{\"u}nchen,            
                   M{\"u}nchen, Germany}                                      
\centerline{$^{25}$Fachbereich Physik, University of Wuppertal,               
                   Wuppertal, Germany}                                        
\centerline{$^{26}$Panjab University, Chandigarh, India}                      
\centerline{$^{27}$Delhi University, Delhi, India}                            
\centerline{$^{28}$Tata Institute of Fundamental Research, Mumbai, India}     
\centerline{$^{29}$University College Dublin, Dublin, Ireland}                
\centerline{$^{30}$Korea Detector Laboratory, Korea University,               
                   Seoul, Korea}                                              
\centerline{$^{31}$SungKyunKwan University, Suwon, Korea}                     
\centerline{$^{32}$CINVESTAV, Mexico City, Mexico}                            
\centerline{$^{33}$FOM-Institute NIKHEF and University of                     
                   Amsterdam/NIKHEF, Amsterdam, The Netherlands}              
\centerline{$^{34}$Radboud University Nijmegen/NIKHEF, Nijmegen, The          
                  Netherlands}                                                
\centerline{$^{35}$Joint Institute for Nuclear Research, Dubna, Russia}       
\centerline{$^{36}$Institute for Theoretical and Experimental Physics,        
                   Moscow, Russia}                                            
\centerline{$^{37}$Moscow State University, Moscow, Russia}                   
\centerline{$^{38}$Institute for High Energy Physics, Protvino, Russia}       
\centerline{$^{39}$Petersburg Nuclear Physics Institute,                      
                   St. Petersburg, Russia}                                    
\centerline{$^{40}$Lund University, Lund, Sweden, Royal Institute of          
                   Technology and Stockholm University, Stockholm,            
                   Sweden, and}                                               
\centerline{Uppsala University, Uppsala, Sweden}                              
\centerline{$^{41}$Physik Institut der Universit{\"a}t Z{\"u}rich,            
                   Z{\"u}rich, Switzerland}                                   
\centerline{$^{42}$Lancaster University, Lancaster, United Kingdom}           
\centerline{$^{43}$Imperial College, London, United Kingdom}                  
\centerline{$^{44}$University of Manchester, Manchester, United Kingdom}      
\centerline{$^{45}$University of Arizona, Tucson, Arizona 85721, USA}         
\centerline{$^{46}$Lawrence Berkeley National Laboratory and University of    
                   California, Berkeley, California 94720, USA}               
\centerline{$^{47}$California State University, Fresno, California 93740, USA}
\centerline{$^{48}$University of California, Riverside, California 92521, USA}
\centerline{$^{49}$Florida State University, Tallahassee, Florida 32306, USA} 
\centerline{$^{50}$Fermi National Accelerator Laboratory,                     
            Batavia, Illinois 60510, USA}                                     
\centerline{$^{51}$University of Illinois at Chicago,                         
            Chicago, Illinois 60607, USA}                                     
\centerline{$^{52}$Northern Illinois University, DeKalb, Illinois 60115, USA} 
\centerline{$^{53}$Northwestern University, Evanston, Illinois 60208, USA}    
\centerline{$^{54}$Indiana University, Bloomington, Indiana 47405, USA}       
\centerline{$^{55}$University of Notre Dame, Notre Dame, Indiana 46556, USA}  
\centerline{$^{56}$Purdue University Calumet, Hammond, Indiana 46323, USA}    
\centerline{$^{57}$Iowa State University, Ames, Iowa 50011, USA}              
\centerline{$^{58}$University of Kansas, Lawrence, Kansas 66045, USA}         
\centerline{$^{59}$Kansas State University, Manhattan, Kansas 66506, USA}     
\centerline{$^{60}$Louisiana Tech University, Ruston, Louisiana 71272, USA}   
\centerline{$^{61}$University of Maryland, College Park, Maryland 20742, USA} 
\centerline{$^{62}$Boston University, Boston, Massachusetts 02215, USA}       
\centerline{$^{63}$Northeastern University, Boston, Massachusetts 02115, USA} 
\centerline{$^{64}$University of Michigan, Ann Arbor, Michigan 48109, USA}    
\centerline{$^{65}$Michigan State University,                                 
            East Lansing, Michigan 48824, USA}                                
\centerline{$^{66}$University of Mississippi,                                 
            University, Mississippi 38677, USA}                               
\centerline{$^{67}$University of Nebraska, Lincoln, Nebraska 68588, USA}      
\centerline{$^{68}$Princeton University, Princeton, New Jersey 08544, USA}    
\centerline{$^{69}$State University of New York, Buffalo, New York 14260, USA}
\centerline{$^{70}$Columbia University, New York, New York 10027, USA}        
\centerline{$^{71}$University of Rochester, Rochester, New York 14627, USA}   
\centerline{$^{72}$State University of New York,                              
            Stony Brook, New York 11794, USA}                                 
\centerline{$^{73}$Brookhaven National Laboratory, Upton, New York 11973, USA}
\centerline{$^{74}$Langston University, Langston, Oklahoma 73050, USA}        
\centerline{$^{75}$University of Oklahoma, Norman, Oklahoma 73019, USA}       
\centerline{$^{76}$Oklahoma State University, Stillwater, Oklahoma 74078, USA}
\centerline{$^{77}$Brown University, Providence, Rhode Island 02912, USA}     
\centerline{$^{78}$University of Texas, Arlington, Texas 76019, USA}          
\centerline{$^{79}$Southern Methodist University, Dallas, Texas 75275, USA}   
\centerline{$^{80}$Rice University, Houston, Texas 77005, USA}                
\centerline{$^{81}$University of Virginia, Charlottesville,                   
            Virginia 22901, USA}                                              
\centerline{$^{82}$University of Washington, Seattle, Washington 98195, USA}  
}                                                                             
\date{January 5, 2007}

\begin{abstract}
  
We have performed the first direct measurement of the time integrated 
flavor untagged charge asymmetry 
in semileptonic $\bs^0$ decays, $\asl$,
 by comparing the decay rate of
$B^{0}_{s} \to \mu^+ D^{-}_{s} \nu X$, where
$D^{-}_{s} \to \phi \pi^{-}$ and
$\phi      \to K^{+} K^{-}$,  with the charge-conjugate $\bsb^0$ decay rate. 
This sample
was selected from 1.3 fb$^{-1}$ of data collected by the D0\ experiment 
in Run II of the
Fermilab Tevatron collider.
We obtain $\asl = [1.23 \pm 0.97\thinspace\mbox{(stat)} \pm 0.17\thinspace\mbox{(syst)}] \times 10^{-2}$.
Assuming that $\Delta m_s / \bar{\Gamma}_s \gg 1$,
this result can be translated into a measurement on the CP-violating phase in $\bs^0$ mixing:
$\Delta \Gamma_s/ \Delta m_s \cdot \tan{\phi_s} = 
 [2.45 \pm 1.93\thinspace\mbox{(stat)} \pm 0.35\thinspace\mbox{(syst)}] \times 10^{-2}$.

\end{abstract}


\maketitle 

This letter presents the first measurement of a time integrated flavor 
untagged charge asymmetry $\asl$ in semileptonic $\bs^0$ decays. 
This asymmetry is defined as:
\begin{equation}
\asl  =  \frac{N(\mu^+ D_s^-) - N(\mu^- D_s^+)}{N(\mu^+ D_s^-) + N(\mu^- D_s^+)},
\end{equation}
where $N(\mu^\pm D_s^\mp)$ is the number of decays 
$\overset{\scriptscriptstyle{(-)}}{B_s^0} \to \mu^\pm D_s^\mp \nu X$ integrated over the $\bs^0$ lifetime.
This asymmetry is called untagged because 
the initial flavor of the $B_s^0$ meson is not determined.
$\asl$ is related to CP violation in $\bs^0$ mixing \cite{pdg}
and can be expressed through the parameters of the $B_s^0$ mass matrix as \cite{nierste}:
\begin{equation}
\asl = \frac{1}{2}\frac{x_s^2 + y_s^2}{1 + x_s^2}\frac{\Delta \Gamma_s}{\Delta m_s}\tan{\phi_s},
\label{theory}
\end{equation}
where $\Delta \Gamma_s$ ($\Delta m_s$) is the width (mass) 
difference between the mass eigenstates in the $\bs^0$ system,
$x_s = \Delta m_s/\bar{\Gamma}_s$, $y_s = \Delta \Gamma_s / (2 \bar{\Gamma}_s)$ where
$\bar{\Gamma}_s$ is the average width in the $\bs^0$ system,
and $\phi_s$ is a CP-violating phase. The standard model (SM) predicts a very small
value for this asymmetry 
$2 \times \asl = a_{SL}^s = (0.21 \pm 0.06) \times 10^{-4}$ \cite{lenz} 
while the contribution
of new physics can significantly modify this prediction \cite{np1, np2}.

This measurement was performed using a large sample of semileptonic 
$\bs^0$ decays
collected by the D0\ experiment at the Fermilab Tevatron collider 
in $p \bar p$ collisions at $\sqrt{s} = 1.96$ TeV
and follows closely the procedure used in the estimate of the dimuon 
asymmetry described in Ref.~\cite{bruce}.
The data correspond to an integrated luminosity of approximately 
1.3 fb$^{-1}$.
The D0 detector is described in detail elsewhere \cite{run2det}.
The detector components most important to this analysis are 
the central tracking
and muon systems. 
The central tracking system consists of a 
silicon microstrip tracker (SMT) and a central fiber tracker (CFT), 
both located within a 2~T superconducting solenoidal 
magnet, with designs optimized for tracking and 
vertexing for pseudorapidities of $|\eta|<3$ and $|\eta|<2.5$, respectively.
The outer muon system, with coverage for $|\eta|<2$, 
consists of a layer of tracking detectors and scintillation trigger 
counters in front of 1.8~T iron toroids, followed by two similar layers 
after the toroids~\cite{run2muon}. The polarities of the solenoid and toroids
are reversed regularly during data taking, so that the four solenoid-toroid
polarity combinations are exposed to approximately the same integrated
luminosity. The direct and reverse magnetic fields in the magnet were
measured to be equal to within 0.1\%. The reversal of magnet polarities
is essential to reduce the detector-related systematics in asymmetry 
measurements
and is fully exploited in this analysis.

The asymmetry $\asl$ was measured using the decay $B^{0}_{s} \to \mu D_{s} \nu X$
with $D_{s} \to \phi \pi$, $\phi      \to K^{+} K^{-}$.
The selection of this final state is described in detail in Ref. \cite{burdin}.
No explicit trigger requirement was made, although most of the sample
was collected with single-muon triggers. Muons were required to have
transverse momentum $p_T(\mu) > 2$ GeV/$c$ and momentum $p(\mu) > 3$ GeV/$c$,
to have hits in both the CFT and SMT, and to have measurements in at least
two layers of the muon system. All reconstructed charged particles in the event
were clustered into jets \cite{durham}, and the $D_s$ candidate was reconstructed
from three tracks found in the same jet as the reconstructed muon.
Oppositely charged particles with $p_T > 0.7$ GeV/$c$ were assigned the kaon
mass and were required to have an invariant mass 
$1.004 < M(K^+ K^-) < 1.034$ GeV/$c^2$,
consistent with that of a $\phi$ meson. The third track was required to have
$p_T > 0.5$ GeV/$c$, a charge opposite to that of the muon charge, and was
assigned the pion mass. The three tracks were required to have hits 
in the CFT
and SMT and to form a common $D_s$ vertex using the algorithm described 
in detail
in \rf{vertex}. 
To reduce combinatorial background, the $D_s$ vertex was required
to have a positive displacement in the transverse plane, relative to 
the $p \bar{p}$
collision point (or primary vertex), with at least 4$\sigma$ significance.
The cosine of the angle between the $D_s$ momentum and the direction from 
the primary
vertex to the $D_s$ vertex was required to be greater than 0.9. 
The trajectories
of the muon and $D_s$ candidates were required to originate from a common
$\bs^0$ vertex, and the ($\mu D_s$) system was required to have an invariant 
mass between 2.6 and 5.4 GeV/$c^2$.

To further improve the $\bs^0$ signal selection, a likelihood ratio 
method \cite{bgv}
was utilized. Using background sidebands $(B)$ and sideband-subtracted 
signal $(S)$
distributions in the data, probability distributions were found for a number
of discriminating variables. These variables were the angle between
the $D_s$ and $K$ momenta in the $K^+K^-$ center-of-mass frame,
the isolation of the $(\mu D_s)$ system, the $\chi^2$ of the $D_s$ vertex, 
the invariant masses $M(\mu D_s)$ and $M(K^+ K^-)$, and $p_T(K^+ K^-)$.
The isolation was defined as the ratio of the sum of the momentum of 
the tracks used to reconstruct the signal divided by the total momentum of
the tracks contained within a cone with 
$\sqrt{\Delta \eta^2 + \Delta \phi^2} < 0.5$ centered on the direction
of the $\mu D_s$ system.
The final requirement on the combined selection likelihood ratio variable
was chosen to maximize the predicted ratio $S/\sqrt{S+B}$.
 
\begin{figure}
\begin{center}
\includegraphics[width=1.0\columnwidth]{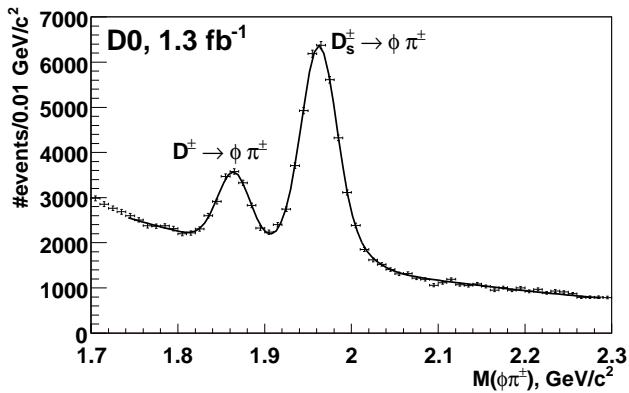}
\caption{The invariant mass  distribution $M(\phi \pi)$ for the selected $\bs^0$ candidates.
The curve shows the result of fit by a function described in the text.}
\label{fig:mkkpi}
\end{center}
\end{figure}

For this analysis we required the $\bs^0$ vertex to have a positive displacement
from the primary vertex to suppress the combinatoric background from the process $c \bar{c} (b \bar{b} ) \to 
\mu D_s \nu X$ with the $D_s$ originating
from a $b$ or $c$ quark, and the muon arising from another quark. 
The invariant mass   distribution $M(\phi \pi)$ for the selected events 
is shown in Fig. \ref{fig:mkkpi}.
The low and high peaks correspond respectively to ($\mu D$),mostly due to 
$B^0$, and ($\mu D_s$), mostly due to $B^0_s$.
The curve represents a fit to the $M(\phi \pi)$ spectrum. 
A single Gaussian was sufficient to describe the
$D \to \phi \pi$ decay, a double Gaussian to describe the
$D_s \to \phi \pi$ decay, and the background was modeled by an exponential.
The total number of events passing all cuts in the $D_s$ mass peak is 
$27,\!300 \pm 300$\thinspace(stat).

To measure $\asl$, both physics and detector effects contributing to the
possible imbalance of events with positively and negatively charged muons must 
be taken into account. One physics source of asymmetry is CP violation
in semileptonic $B$ decays. In addition, forward-backward charge asymmetry
of events produced in the proton-antiproton collisions can also be present.
Detector effects can give rise to an artificial asymmetry if, e.g., the reconstruction
efficiencies of positively and negatively charged particles are different. 
However, a positively charged 
particle produces the same track as a negatively charged particle in the detector
with reversed magnet polarity. Therefore, almost all detector effects can be canceled
provided the fractions of events with opposite magnet polarities are approximately
the same. This is the case in this analysis, where the exposures are the
same within 1\%.

According to the method described in \rf{bruce},
the event sample was divided into eight subsamples corresponding to 
all possible combinations of the toroid polarity $\beta = \pm 1$, 
the sign   of the  pseudorapidity of the ($\mu \phi \pi$) system \cite{pseudo}
$\gamma = \pm 1$, and the sign of the muon charge $q = \pm 1$.
The number of $(\mu D_s)$ events in each subsample was obtained by a fit to the
mass distribution $M(\phi \pi)$ using the same function as for the whole sample.
For the cross-check we also extracted the numbers of $(\mu D)$ and background events from the fit.
The widths and positions of the $(\mu D_s)$ and $(\mu D)$ peaks, 
the relative fractions of the two Gaussians describing the $(\mu D_s)$ peak,
as well as the background slope were fixed to the values obtained from the fit to the total 
$M(\phi \pi)$ distribution. The numbers of ($\mu D_s$) and
($\mu D$) events, $n^{\beta \gamma}_q(D_s)$ and $n^{\beta \gamma}_q(D)$, 
along with the number of the background events in
the fitting range $1.75 - 2.30$ GeV$/c^2$, $n^{\beta \gamma}_q(\mbox{bkg})$,
for each subsample is given in Table \ref{tab:bgq}.

The fitted numbers of ($\mu D_s$) [($\mu D$), background] events were used to disentangle 
the physics asymmetries and the detector
effects. The $n^{\beta \gamma}_q$ can be expressed through the physics and the detector 
asymmetries as follows \cite{bruce}:
\begin{eqnarray}
n_{q}^{\beta \gamma} & = & \frac{1}{4} N \epsilon^{\beta}(1+qA)
(1+q \gamma A_{\rm fb})
(1+\gamma A_{\rm det}) \nonumber \\
                     & \times & (1+q \beta \gamma A_{\rm ro}) 
(1+q \beta A_{q \beta}) (1+\beta \gamma A_{\beta\gamma}).
\label{detpar}
\end{eqnarray}
Here
$N$                   is the total number of ($\mu D_s$) [($\mu D$), background] events;
$\epsilon^{\beta}$    is the fraction of integrated luminosity with toroid polarity $\beta$ 
                      ($\epsilon^+ + \epsilon^- = 1$);
$A$                   is the integrated charge asymmetry to be measured;
$A_{\rm fb}$              is the forward-backward asymmetry;
$A_{\rm det}$             is the detector asymmetry for particles emitted in the forward 
                      and backward direction;
$A_{\rm ro}$              is the range-out asymmetry 
that accounts for the change
                      in acceptance 
                      of muons which bend towards the beam line and those
which bend away from the beam line;
$A_{q \beta}$         is the detector asymmetry which accounts for the change in
                      the muon reconstruction efficiency when the toroid polarity is reversed;
$A_{\beta \gamma}$    accounts for any detector related forward-backward asymmetries that remain after the toroid polarity flip.

\begin{table}
\begin{center}
\caption{The numbers of events $n^{\beta \gamma}_q(D_s)$ [$n^{\beta \gamma}_q(D)$]  
in the $D_s$ [$D$] mass peak and in the background $n^{\beta \gamma}_q(\mbox{bkg})$ 
for eight subsamples.} 
\begin{ruledtabular}
\begin{tabular}{cD{.}{\,\pm\,}{-1}D{.}{\,\pm\,}{-1}D{.}{\,\pm\,}{-1}}
Subsample:            & \multicolumn{1}{c}{$n^{\beta \gamma}_q(D_s)$} & \multicolumn{1}{c}{$n^{\beta \gamma}_q(D)$} & \multicolumn{1}{c}{$n^{\beta \gamma}_q(\mbox{bkg})$} \\
  $ \beta \gamma q$   &   \multicolumn{1}{c}{(events)}             &      \multicolumn{1}{c}{(events)}        & \multicolumn{1}{c}{(events)} \\
\hline
$+++$ & 3,\!216 . 76 & 907   . 55   & 9,\!797  . 124 \\
$+-+$ & 3,\!586 . 79 & 965   . 56   & 10,\!387 . 127 \\
$++-$ & 3,\!391 . 78 & 1,\!037 . 57   & 10,\!390 . 127 \\
$+--$ & 3,\!225 . 76 & 963   . 55   & 9,\!832  . 124 \\
$-++$ & 3,\!616 . 80 & 1,\!003 . 57   & 10,\!508 . 128 \\
$--+$ & 3,\!370 . 77 & 801   . 54   & 9,\!987  . 125 \\
$-+-$ & 3,\!353 . 77 & 831   . 55   & 10,\!215 . 125 \\
$---$ & 3,\!532 . 79 & 1,\!116 . 59   & 10,\!701 . 129 \\
\end{tabular}
\end{ruledtabular}
\label{tab:bgq}
\end{center}
\end{table}

Since the system (\ref{detpar}) contains eight equations, all six asymmetries together with
$N$ and $\epsilon^+$ can be extracted for each of the three types of the events.
Results are presented in Table \ref{tab:res} separately for ($\mu D_s$) and ($\mu D$) events and the background.
The physics asymmetries $A$ and $A_{\rm fb}$ for background events 
are consistent with zero. This is an important
test for this method, since the precision of the asymmetry measurement 
for the background events is much higher than that of the signal due to the larger statistics. 
The largest detector asymmetry for all three types of the events is
the range-out asymmetry. 

It can be seen from (\ref{detpar}) that if 
$\epsilon^+ = \epsilon^- = 1/2$, and the asymmetries $A$, $A_{\rm fb}$, 
$A_{\rm det}$, $A_{\rm ro}$, $A_{q\beta}$, $A_{\beta \gamma}$ are small,
each of them can be obtained
independently by the appropriate division of 
the entire sample of events into two parts. For example, the asymmetry $A$
can be obtained be dividing the sample according to
the charge of muon. For such a division, and neglecting the second
order terms, we obtain:
\begin{eqnarray}
A & = & \frac{n_+ - n_-}{n_+ + n_-},  \\
n_q & = & \sum_{\beta, \gamma = -1}^{+1} n_q^{\beta \gamma} 
\simeq \frac{1}{2}N (1 + q A). \nonumber
\end{eqnarray}
This observation explains in particular 
the similar values of statistical uncertainties for all asymmetries
in Table \ref{tab:res}.


\begin{table}
\begin{center}
\caption{The physics and detector asymmetries for ($\mu D_s$), 
($\mu D$) and background events. Uncertainties are statistical.}
\begin{ruledtabular}
\begin{tabular}{cD{+}{\,\pm\,}{-1}D{+}{\,\pm\,}{-1}D{+}{\,\pm\,}{-1}}

 & \multicolumn{1}{c}{$(\mu D_s)$} & \multicolumn{1}{c}{($\mu D$)} & \multicolumn{1}{c}{Background} \\

\hline
$N$            & 27,\!289 + 220      & 7,\!623 + 162      & 81,\!817 + 357   \\
$\epsilon^{+}$ & 0.492  + 0.004    & 0.510 + 0.011    & 0.494  + 0.002 \\
\hline 
$A$                  &  0.0102 + 0.0081 & -0.0345 + 0.0211 & -0.0056 + 0.0045  \\
$ A_{\rm fb}$            & -0.0046 + 0.0081 &  0.0480 + 0.0210 & -0.0020 + 0.0043  \\
$A_{\rm det}$            & -0.0051 + 0.0081 & -0.0072 + 0.0212 &  0.0001 + 0.0044  \\
$ A_{\rm ro}$            & -0.0352 + 0.0081 & -0.0819 + 0.0209 & -0.0263 + 0.0044  \\
$ A_{\beta \gamma}$  & -0.0097 + 0.0081 &  0.0104 + 0.0213 & -0.0010 + 0.0044  \\
$ A_{q \beta}$       &  0.0030 + 0.0081 &  0.0014 + 0.0212 &  0.0046 + 0.0044  \\
\end{tabular}
\end{ruledtabular}
\label{tab:res}
\end{center}
\end{table}

The resulting charge asymmetry of ($\mu D_s$) events is
$A = 0.0102 \pm 0.0081\thinspace\mbox{(stat)}$. It is related to $\asl$
via \mbox{$A = f_s \cdot \asl + f_d \cdot \asld$}, 
where $f_s$($f_d$) is the fraction of $\bs^0(B_d^0) \to \mu D_s \nu X$ decays in the ($\mu D_s$) sample.
$\asld$ may arise only from $B_d^0 \to D D_s$ decay, 
the fraction of which in the ($\mu D_s$) sample was found to be small, at the level
of $(4 \pm 1)$\%.
Additionally, the value of $\asld$ is strongly constrained experimentally \cite{babar, belle} to be close to zero.
Therefore the time-integrated $\asld$ component can be neglected.
The fraction of $\bs^0$ decays, $f_s$, was determined as follows.
The decays $B_s^0 \to \mu D_s \nu X$ and $B_s^0 \to \tau D_s \nu X \to \mu D_s \nu X$
were considered as a signal. The decays $B_s^0 \to D_s D_s X$ with 
$D_s \to \mu \nu X$ are not flavor-specific and hence were considered
as a background. The decays $B_d^0 \to D D_s X$ were also included in the background.
In addition, the process $c \bar{c} (b \bar{b} ) \to 
\mu D_s \nu X$ was taken into account. This background produces a pseudovertex
which peaks around the primary interaction point. It is reduced by approximately 50\%
by requiring a positive displacement of the ($\mu D_s$) vertex.

All processes were simulated using the 
{\sc evtgen} \cite{EvtGen} generator interfaced 
to {\sc pythia} \cite{pyth} and followed 
by full modeling of the detector response using {\sc geant} \cite{geant} and event reconstruction.
The branching fractions of $B_d^0$ decays were taken from \rf{pdg},
while the contribution of the process $c \bar{c} (b \bar{b} ) \to 
\mu D_s \nu X$ was measured directly in our data to be $(5.9 \pm 1.7)\%$.
With these assumptions,  $(83.2 \pm 3.3)$\% of the selected sample
of ($\mu D_s$) events is composed of semileptonic $\bs^0$
decays. The uncertainty on this value comes from the uncertainties on the branching
ratios of the contributing $B$ decays and the uncertainty on the fraction of the 
$c \bar{c} (b \bar{b} ) \to \mu D_s \nu X$ process in the sample.
Taking into account the sample composition, 
the measured integrated charge asymmetry of semileptonic $\bs^0$
decay is found to be $\asl = [1.23 \pm 0.97\thinspace\mbox{(stat)}] \times 10^{-2}$.

The following sources of systematic uncertainty were considered.
The final state includes a $K^+ K^-$ pair. Therefore, the charge
asymmetry of $K$ meson reconstruction, which arises due to the different
interaction cross sections of $K^+$ and $K^-$ in the detector material, 
does not contribute to the measured $\asl$. The charge asymmetry of pion
reconstruction, however, can contribute. The  $\pi d$ interaction cross sections
for positive and negative pions differ by $(1.3 \pm 0.3)\%$ in the range
$1-2$ GeV$/c$ \cite{carter}. Taking into account the amount of material which a pion
crosses in the detector, the induced asymmetry due to pion reconstruction
was estimated to be $2 \times 10^{-4}$. This value was included in the systematic uncertainty.

The uncertainty in the fraction of $\bs^0$ signal in 
the ($\mu D_s$) sample produces a systematic uncertainty of $1 \times 10^{-3}$. 
This uncertainty also includes a possible 
residual variation of the signal fraction
between subsamples. 

The uncertainty due to the fitting procedure was estimated by varying the masses
and widths of the peaks, and the slope of the background by one standard deviation. The fitting
procedure was also repeated with a single Gaussian describing the $D_s$ peak 
and with a different fitting range. The resulting change of $\asl$ did not exceed
$0.14 \times 10^{-2}$ which was used as an estimate of the systematic uncertainty
from this source. 

The $\bs^0$ reconstruction efficiency varies with the decay length due to the applied requirements.
We verified that this variation does not bias the result for $\asl$ and the relation (\ref{theory}). 
In addition, any possible contribution of the $B_d^0$ charge asymmetry to the measured
value was estimated to be negligible. 

Adding all contributions into the systematic uncertainty in quadratures, we
obtain the resulting value of the time-integrated untagged charge asymmetry:
\begin{equation}
\asl = [1.23 \pm 0.97\thinspace\mbox{(stat)} \pm 0.17\thinspace\mbox{(syst)}] \times 10^{-2}.
\end{equation}

%
%


This is the first direct measurement of $\asl$. It can be seen that the statistical
uncertainty dominates and will be improved in the future 
with the increase of statistics and addition of new decay modes.
Using Eq.\ (\ref{theory}) and assuming that $\Delta m_s / \bar{\Gamma}_s \gg 1$,
we obtain:

\begin{eqnarray}
& & \frac{\Delta \Gamma_s}{\Delta m_s}\tan{\phi_s} = \nonumber \\
& & [2.45 \pm 1.93\thinspace\mbox{(stat)} \pm 0.35\thinspace\mbox{(syst)}] \times 10^{-2}.
\end{eqnarray}
This result, together with the measurements of $\Delta \Gamma_s$ \cite{dgam-new}
and $\Delta m_s$ \cite{burdin,cdf}, provides a constraint on the CP-violating 
phase $\phi_s$.

%
We thank the staffs at Fermilab and collaborating institutions, 
and acknowledge support from the 
DOE and NSF (USA);
CEA and CNRS/IN2P3 (France);
FASI, Rosatom and RFBR (Russia);
CAPES, CNPq, FAPERJ, FAPESP and FUNDUNESP (Brazil);
DAE and DST (India);
Colciencias (Colombia);
CONACyT (Mexico);
KRF and KOSEF (Korea);
CONICET and UBACyT (Argentina);
FOM (The Netherlands);
PPARC (United Kingdom);
MSMT (Czech Republic);
CRC Program, CFI, NSERC and WestGrid Project (Canada);
BMBF and DFG (Germany);
SFI (Ireland);
The Swedish Research Council (Sweden);
Research Corporation;
Alexander von Humboldt Foundation;
and the Marie Curie Program.
%



\begin{thebibliography}{99}
\bibitem{pseudo}
Pseudorapidity $\eta$ is defined as $\eta = \ln (\tan (\theta/2))$ where
$theta$ is the polar angle of particle with respect to the beam direction.
\end{thebibliography}
\end{document}